\documentstyle[epsfig,rotate]{lamuphys}
\makeatletter
\let\chapter\hid@chapter
\makeatother
\begin{document}
\pagenumbering{arabic}
\title{From$\!$ Microscopic$\!$ to$\!$ Macroscopic$\!$ Traffic$\!$ Models$\!$}
 
\author{Dirk Helbing\inst{1,2}}
 
\institute{II. Institute of Theoretical Physics, University of Stuttgart,
Pfaffenwaldring 57/III, 70550 Stuttgart, Germany
\and
Department of Fluid Mechanics and Heat Transfer,
Tel Aviv University, Tel Aviv, 69978, Israel}
 
\maketitle
 
\begin{abstract}
The paper presents a systematic derivation of macroscopic equations
for freeway traffic flow from an Enskog-like kinetic approach.
The resulting fluid-dynamic traffic equations for the
spatial density, average velocity, and velocity variance of
vehicles are compared to equations, which can be obtained 
from a microscopic force model of individual vehicle motion.
Simulation results of the models are confronted with empirical traffic data.
\end{abstract}
\section{Introduction}

During the last five years, modelling and simulating traffic dynamics
has found a large and rapidly growing interest in physics. This is due to
\begin{figure}[htbp]
\unitlength7mm
\begin{center}
\begin{picture}(16,9.8)(0,0)
\put(0,9.8){\epsfig{height=16\unitlength, width=9.8\unitlength, angle=-90, 
      bbllx=50pt, bblly=50pt, bburx=554pt, bbury=770pt, 
      file=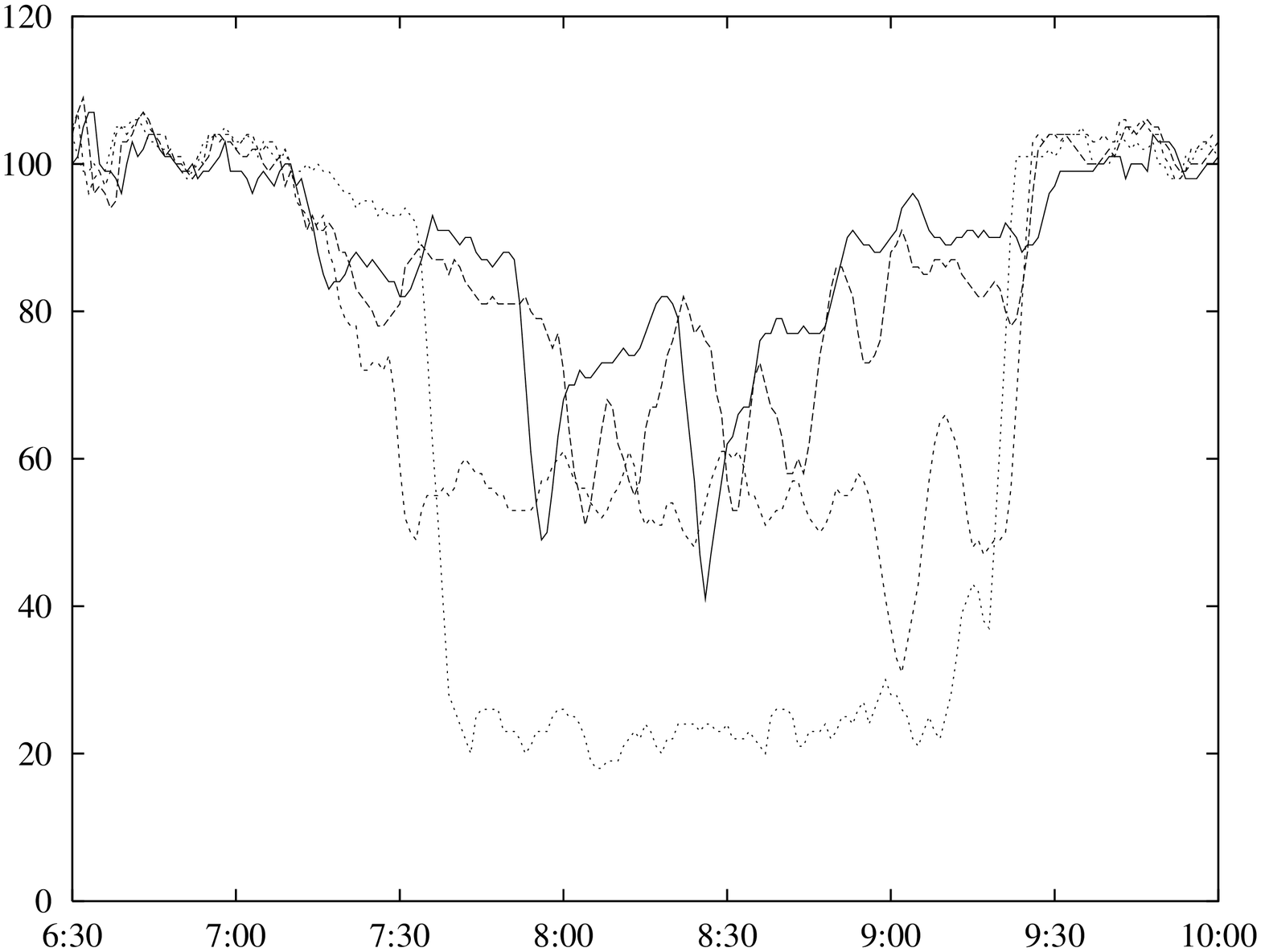}}
\put(8.7,-0.6){\makebox(0,0){\small $t$ (h)}}
\put(0.2,5.1){\makebox(0,0){\rotate[l]{\hbox{\small $V(r,t)$ (km/h)}}}}
\end{picture}
\end{center}
\caption[]{Temporal evolution of the average velocity $V(r,t)$ at subsequent
cross-sections of the Dutch highway A9 from Haarlem to Amsterdam
at October 14, 1994 (five minute averages of single vehicle data)
\cite{empi,VD}. The prescribed speed limit is 120\,km/h. We observe
a breakdown of velocity during the rush hours between 7:30\,am and
9:30\,am due to the overloading of the highway at
$r = r_0 := 41.8$\,km ($\cdots$). 
At the subsequent cross-sections the traffic situation recovers 
(-~-~-: $r = r_0+1$\,km; --~--: $r = r_0+2.2$\,km; ---: $r = r_0+4.2$\,km).
Nevertheless, the amplitudes of the small velocity fluctuations at $r_0$
become larger and larger, leading to so-called stop-and-go waves.} 
\label{inst}
\end{figure}
\begin{enumerate}
\item similarities of traffic dynamics with flows of gases, fluids, 
and granular media,
\item instability phenomena and critical behavior of traffic (cf. Fig.
\ref{inst}),
\item interesting applications of cellular automata and molecular dynamics
simulation methods,
\item the need of efficient traffic optimization methods in order to keep
or increase the level of mobility. 
\end{enumerate}
Usually, one distinguishes three levels of modelling: The microscopic level
of description delineates the dynamics of the single driver-vehicle units
\cite{follow1,follow2,Wiede,Bando,VD}.
This allows to consider different vehicle characteristics and
driving styles, so that many of these models aim at a high-fidelity 
description of traffic flow, e.g. \cite{Wiede,VD}. 
They are mostly used for detail studies 
(e.g. of on-ramp traffic, bottlenecks, effects of traffic
optimization measures), but they consume an enourmous amount of CPU time
due the the large number of variables involved. 
An alternative approach are cellular automata, 
which allow to simulate
a minimal model of traffic dynamics faster than real-time 
\cite{Cell1,Cell2,Cell3,Cell4}. 
\par
Computational efficiency can also be reached by macroscopic traffic 
models, but at a higher degree of accuracy \cite{Payne,KK1,KK2,PhysA,VD}.
Macroscopic traffic models consist
of equations for a few aggregate quantities like the spatial density $\rho$,
the average velocity $V$, and (in some cases) additional velocity moments.
These equations are similar to fluid-dynamic equations, but 
some fundamental differences with respect to the dynamics of ordinary 
fluids have recently been recognized \cite{VD,PhysA}. For congested conditions,
their detailled form is not at all obvious. Therefore, it has been suggested
to derive the macroscopic traffic equations from a kinetic, i.e. mesoscopic
level of description, which delineates the spatio-temporal evolution of
the velocity distribution \cite{VD,PhysA,Kin1,Kin2,Kin3,Kin4}.

\section{Enskog-like Kinetic Traffic Model}

In the following, we define the coarse-grained phase-space density
$\tilde{\rho}(r,v,t)$ of vehicles per lane with velocity $v$ at place $r$ and 
time $t$ by
\begin{equation}
 \tilde{\rho}(r,v,t) := \frac{1}{(2\,\Delta r)(2\, \Delta v)} \sum_\alpha
 \!\!\! \int\limits_{r-\Delta r}^{r+\Delta r} \!\!\! dr' 
 \!\!\! \int\limits_{v-\Delta v}^{v+\Delta
   v} \!\!\! dv' \, \delta(r' - r_\alpha(t)) \, \delta (v' - v_\alpha(t)) \, .
\label{phase}
\end{equation}
$r_\alpha(t)$ is the location and $v_\alpha(t)$ the velocity of vehicle
$\alpha$ at time $t$. We do not distinguish
different lanes, here, although this is possible \cite{VD,Lanes1}. 
Instead, we treat the overall cross section of an
$n$-lane freeway in an effective way \cite{VD,Lanes2}.
\par
Let us assume an acceleration equation of the form
\begin{equation}
 \frac{dv_\alpha}{dt} = f_0(v_\alpha)
 + \sum_{\beta(\ne \alpha)}
 f_{\alpha\beta}(r_\alpha,v_\alpha,r_\beta,v_\beta) + \xi_\alpha(t) \, ,
\label{force}
\end{equation}
where the function
\begin{equation}
 f_0(v_\alpha) := \frac{V_0-v_\alpha(t)}{\tau}
\end{equation}
describes an adaptation of the actual velocity 
$v_\alpha(t)$ to the desired velocity $V_0$ within a  
(possibly density-dependent)
relaxation time $\tau$. The second term in (\ref{force})
delineates the effect of interactions with vehicles $\beta$, and $\xi_\alpha(t)$
reflects velocity fluctuations due to imperfect driving. We will assume
$\langle \xi_\alpha(t) \xi_\beta(t') \rangle 
= 2D \, \delta_{\alpha\beta} \,
\delta(t-t')$, where the diffusion function $D$ is density- and
velocity-dependent
\cite{VD,PhysA,Lett}. For reasons of simplicity, the desired velocity 
$V_0$ and the relaxation time $\tau$ were taken identical for all
vehicles, but it is also possible to generalize this model 
\cite{Kin2,Kin3,Kin4}.
\par
From (\ref{phase}) and (\ref{force}), the
following dynamical equation for the phase-space density can be derived:
\begin{equation}
 \frac{\partial \tilde{\rho}}{\partial t}
 + \frac{\partial (\tilde{\rho} v)}{\partial r}
 + \frac{\partial}{\partial v} \left[\tilde{\rho} \, f_0(v) 
 \right] = \left( \frac{\partial \tilde{\rho}}{\partial t} \right)_{\rm fl}
 + \left( \frac{\partial \tilde{\rho}}{\partial t} \right)_{\rm int} \, .
\end{equation}
The fluctuation term gives a contribution
\begin{equation}
 \left( \frac{\partial \tilde{\rho}}{\partial t} \right)_{\rm fl}
 = \frac{\partial^2 (\tilde{\rho} D)}{\partial v^2} \, .
\end{equation}
In addition, we have used the abbreviation
\begin{equation}
 \left( \frac{\partial \tilde{\rho}}{\partial t} \right)_{\rm int}
 := - \frac{\partial}{\partial v} (\tilde{\rho} f_{\rm int})
\label{INT}
\end{equation}
with the average interaction force
\begin{equation}
 f_{\rm int}(r,v,t) := \frac{1}{4 \tilde{\rho} \, \Delta r \, \Delta v}
 \sum_{\alpha \ne \beta} 
 \int\limits_{r-\Delta r}^{r+\Delta r} \!\!\! dr' 
 \!\!\! \int\limits_{v-\Delta v}^{v+\Delta
   v} \!\!\! dv' \, f_{\alpha\beta} \, 
 \delta(r' - r_\alpha(t)) \, \delta (v' - v_\alpha(t)) \, .
\end{equation}
The interaction term (\ref{INT}) reflects deceleration processes. In analogy to
the Enskog theory of dense gases \cite{chapman} and granular media
\cite{densegran1,densegran2,gold}, 
but with an interaction law 
typical for vehicles \cite{VD,PhysA}, it is approximated by 
\begin{equation}
 \left( \frac{\partial \tilde{\rho}}{\partial t} \right)_{\rm int} 
 = (1-p) \chi(r+l,t) {\cal B}(v) 
\end{equation}
with the Boltzmann-like interaction function
\begin{eqnarray}
 {\cal B}(v) &=& \!\!\int\limits_{w>v} \!\! dw \, (w-v) \, \tilde{\rho}(r,w,t)
 \tilde{\rho}(r+s,v,t) \nonumber \\
 &-& \!\!\int\limits_{v>w}\!\! dw \, (v-w) 
 \tilde{\rho}(r,v,t)\tilde{\rho}(r+s,w,t) \, .
\label{Boltz}
\end{eqnarray}
According to this, the phase-space density $\tilde{\rho}(r,v,t)$ increases
due to deceleration of vehicles with velocities $w>v$, which cannot 
overtake vehicles with velocity $v$. The density-dependent
probability of immediate overtaking is repre\-sented by $p$.
A decrease of the phase-space density $\tilde{\rho}(r,v,t)$ is caused by
interactions of vehicles with velocity $v$ with slower vehicles driving
with velocities $w<v$. The corresponding interaction rates are proportional
to the relative velocity $|v-w|$ and to the phase space densities 
of both interacting vehicles. By $s(V) = l_0 + l(V)$ ($\approx$ 
vehicle length $+$ safe distance) it is taken into account
that the distance of interacting vehicles is given by
their velocity-dependent space requirements.
These cause an increase of the interaction rate, which is described by the
pair correlation function $\chi(r) = [1- \rho(r,t) s]^{-1}$
at the 'interaction point' $r+l$.
A more detailled discussion of the above kinetic 
traffic model is presented elsewhere \cite{VD,PhysA}. 
\par
\begin{figure}[htbp]
\unitlength7mm
\begin{center}
\begin{picture}(16,9.8)(0,0)
\put(0,9.8){\epsfig{height=16\unitlength, width=9.8\unitlength, angle=-90, 
      bbllx=50pt, bblly=50pt, bburx=554pt, bbury=770pt, 
      file=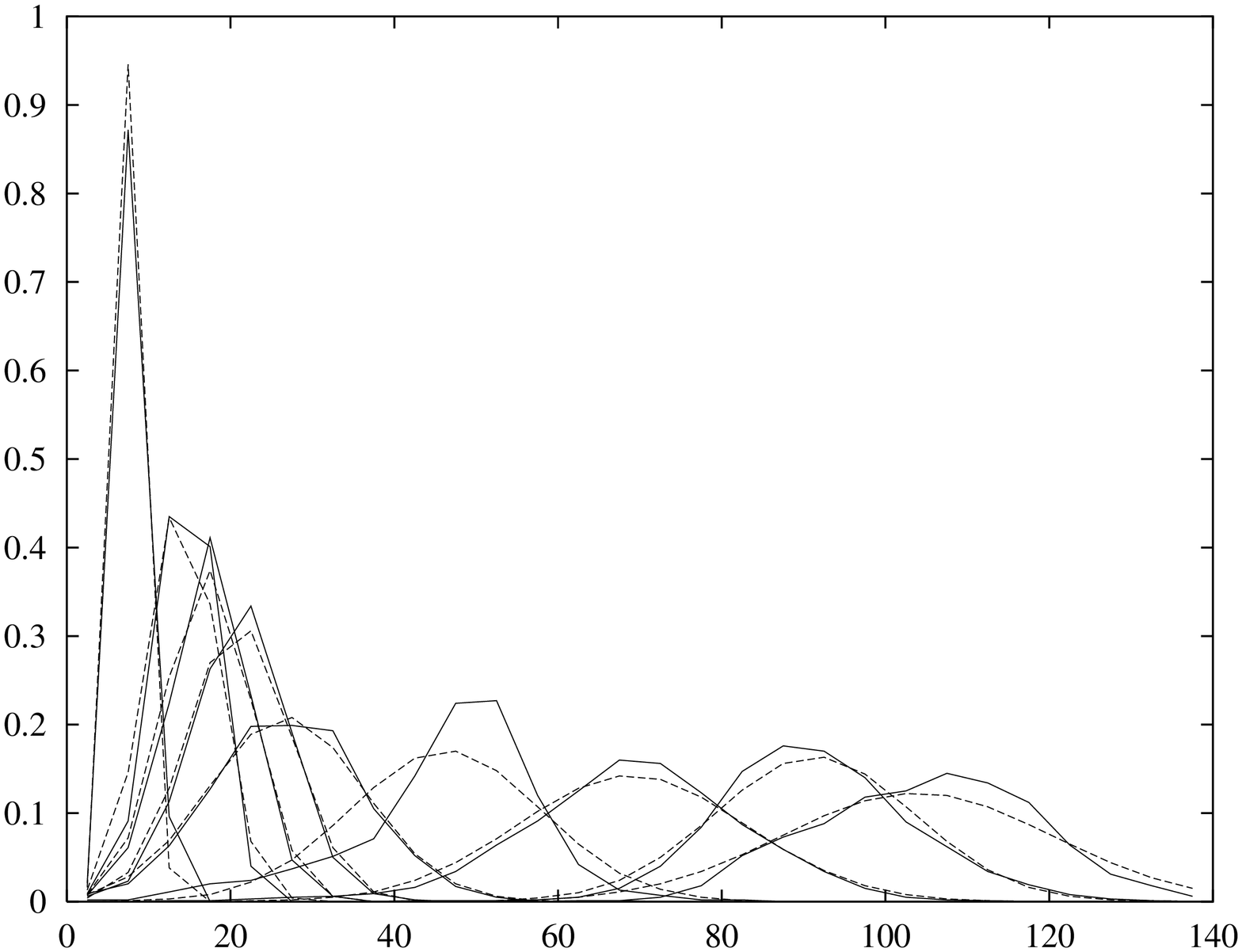}}
\put(8.7,-0.6){\makebox(0,0){\footnotesize $v$ (km/h)}}
\put(0.2,5.1){\makebox(0,0){\rotate[l]{\hbox{\footnotesize $P(v;r,t)$}}}}
\put(5.4,8){\makebox(0,0){\footnotesize $\rho=90$\,vehicles/km lane}}
\put(3.43,4.7){\makebox(0,0){\footnotesize 80}}
\put(4.24,4){\makebox(0,0){\footnotesize 70}}
\put(4.69,3.2){\makebox(0,0){\footnotesize 60}}
\put(5.15,2.5){\makebox(0,0){\footnotesize 50}}
\put(7.44,2.5){\makebox(0,0){\footnotesize 40}}
\put(8.76,2.5){\makebox(0,0){\footnotesize 30}}
\put(10.64,2.5){\makebox(0,0){\footnotesize 20}}
\put(12.35,2.2){\makebox(0,0){\footnotesize 10}}
\end{picture}
\end{center}
\caption[]{Comparison of empirical velocity distributions at different
densities (---) with frequency polygons of 
grouped Gaussian velocity distributions with
the same mean value and variance (--~--) \cite{emp}. 
A significant deviation of the empirical relations from the
respective discrete Gaussian approximations is only found at a density
of $\rho = 40$\,vehicles/km lane, where the temporal averaging period of
$T=2$\,min may have been too long due to rapid stop-and-go waves.}
\label{F2}
\end{figure}
\begin{figure}[htbp]
\unitlength7mm
\begin{center}
\begin{picture}(16,9.6)(0,0)
\put(-0.2,9.8){\epsfig{height=16\unitlength, width=9.8\unitlength, angle=-90, 
      bbllx=50pt, bblly=50pt, bburx=554pt, bbury=770pt, 
      file=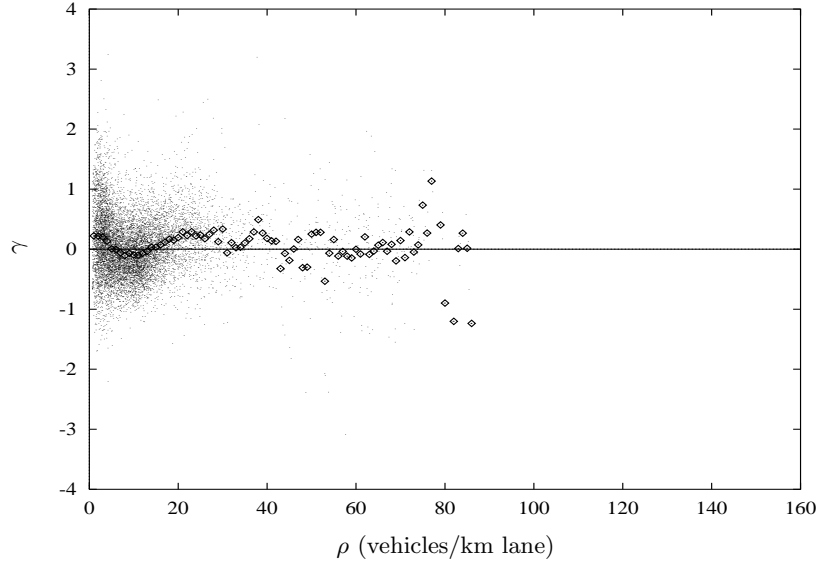}}
\put(8.5,-0.6){\makebox(0,0){\footnotesize $\rho$ (vehicles/km lane)}}
\put(0.4,5.1){\makebox(0,0){\rotate[l]{\hbox{\footnotesize $\gamma$}}}}
\end{picture}
\end{center}
\caption[]{Density-dependence of the skewness $\gamma$ of the
velocity distribution
($\cdot$\,: 1-minute data; $\Diamond$: respective mean values) \cite{emp}.
The large variation of the 1-minute data at low densities is due to the
small number of vehicles which pass a cross section during the time
interval $T=1$\,min, whereas the large variation of their mean values
at high densities comes from the few 1-minute data, over
which could be averaged. The 1-minute data of the skewness scatter around
the zero line (---) and mostly lie between $-1$ and $1$, so that it
is negligible most of the time.} 
\label{F4}
\end{figure}

Now, we will focus on the the macroscopic equations for the spatial density
\begin{equation} 
 \rho(r,t) = \int dv \, \tilde{\rho}(r,v,t) \, , 
\end{equation}
the average velocity
\begin{equation}
 V(r,t) = \int dv \, v \frac{\tilde{\rho}(r,v,t)}{\rho(r,t)} \, , 
\end{equation}
and the velocity variance
\begin{equation}
 \theta(r,t) = \int dv \, [v-V(r,t)]^2 \frac{\tilde{\rho}(r,v,t)}
 {\rho(r,t)} \, . 
\end{equation}
These are obtained by multiplying the kinetic equation with $v^k$ and integrating
with respect to $v$. In order to obtain a closed system of equations,
we assume that the velocity distribution $P(v;r,t)$ has a Gaussian form, i.e.
\begin{equation}
 P(v;r,t) := \frac{\tilde{\rho}(r,v,t)}{\rho(r,t)} 
 = \frac{\mbox{e}^{-[v-V(r,t)]^2/[2\theta(r,t)]}}{\sqrt{2\pi \theta(r,t)}} \, .
\end{equation}
According to empirical data, this approximation is well justified
(cf. Figs. \ref{F2} and \ref{F4}). 
\par
After some straightforward calculations,
the following equations are obtained:
\begin{equation}
 \frac{\partial \rho}{\partial t} + \frac{\partial (\rho V)}{\partial r}
 = 0 \, ,
\label{contin}
\end{equation}
\begin{equation}
 \frac{\partial (\rho V)}{\partial t} + \frac{\partial}{\partial r}
 [\rho (V^2 + \theta)] = \frac{\rho}{\tau} ( V_0 - V ) 
 + (1-p) \chi(r+l,t) \int dv \, v {\cal B}(v) \, ,
\label{veloc}
\end{equation}
\begin{eqnarray}
 \frac{\partial}{\partial t} [\rho (V^2 + \theta)] +
 \frac{\partial}{\partial r} [\rho (V^3 + 3 V \theta )]
 &=& \frac{2\rho}{\tau} (V_0V + \tau D - V^2 - \theta) \nonumber \\
 &+& (1-p)\chi(r+l,t)
 \int dv\, v^2 {\cal B}(v) \, . \qquad
\label{var}
\end{eqnarray}
Equations (\ref{contin}) to (\ref{var}) are similar to the Euler equations
of ordinary fluids. In particular, the density equation
(\ref{contin}) agrees with the continuity equation, 
reflecting that the number of vehicles is conserved (on a circular road).
However, equations (\ref{veloc}) and (\ref{var}) contain some additional
terms compared to the hydrodyamic equations for momentum density and
energy density, which are essential for the instability of traffic flow. 
The respective first terms on the right-hand sides of
(\ref{veloc}) and (\ref{var}) originate from the acceleration
towards the desired velocity $V_0$ and from velocity fluctuations.
The respective last terms reflect interaction (deceleration) effects.
In contrast to ordinary fluids, they do not vanish, since vehicular
interactions do not conserve momentum and energy.
\par
To obtain the explicit form of the interaction terms, one has to carry out
a number of lengthy calculations. 
Using the abbreviations
\begin{equation}
 \rho_+(r,t) := \rho(r+s,t) \, , \quad V_+(r,t) := V(r+s,t) \, , \quad 
 \theta_+(r,t) := \theta(r+s,t) \, ,
\end{equation}
and introducing the Gaussian error function
\begin{equation}
 \Phi(x) = \int\limits_{-\infty}^x dy \, \frac{\mbox{e}^{-y^2/2}}{\sqrt{2\pi}}
 \, ,
\end{equation}
one finally finds
\begin{eqnarray}
 \int dv \, v {\cal B}(v) &=& - \rho \rho_+ \left\{ \Big[ (\theta + \theta_+)
 + (V - V_+)^2 \Big]
 \, \Phi \left( \frac{V-V_+}{\sqrt{\theta+\theta_+}} \right) \right.
 \nonumber \\
 & & 
 + \left. (V-V_+)(\theta+\theta_+) 
 \frac{\mbox{e}^{-(V-V_+)^2/[2(\theta+\theta_+)]}}{\sqrt{2\pi(\theta+\theta_+)}}
 \right\} \label{eins}
\end{eqnarray}
and
\begin{eqnarray}
 \int dv \, v^2 {\cal B}(v) &=& 
 - 2 \rho \rho_+ (\theta-\theta_+) \left[ (\theta+\theta_+) 
 \frac{\mbox{e}^{-(V-V_+)^2/[2(\theta+\theta_+)]}}{\sqrt{2\pi(\theta+\theta_+)}}
 \right. \nonumber \\
 & & + (V-V_+) \left. \Phi \left( 
 \frac{V-V_+}{\sqrt{\theta+\theta_+}} \right) \right] 
 + (V+V_+) \int dv \, v {\cal B}(v) \, . \qquad
\label{zwei}
\end{eqnarray}
The macroscopic traffic equations (\ref{contin}) to (\ref{var}) were 
written as equations for fluxes with sink/source terms (the terms on the
right-hand side). The flux representation is very advantageous, since
many numerical integration algorithms have been developed for this 
type of partial differential equations. However, due to (\ref{eins}) and
(\ref{zwei}), the flux representation is non-local. This is caused by the
finite space requirements of cars, i.e. a driver reacts to another car 
already at a certain distance. As a consequence, the non-local interaction
terms imply viscosity effects, among other things. To see this, we 
expand them up to second order. Neglecting products of spatial derivatives,
we get the continuity equation
\begin{equation}
 \frac{\partial \rho}{\partial t} + V \frac{\partial \rho}{\partial r}
 = - \rho \frac{\partial V}{\partial r} \, ,
\label{dens}
\end{equation}
the velocity equation
\begin{eqnarray}
 \frac{\partial V}{\partial t} + V \frac{\partial V}{\partial r} 
 &=& a_1 \frac{\partial \rho}{\partial r} 
 + a_2 \frac{\partial V}{\partial r}
 + a_3 \frac{\partial \theta}{\partial r} \nonumber \\
 &+& b_1 \frac{\partial^2 \rho}{\partial r^2} 
  + b_2 \frac{\partial^2 V}{\partial r^2}
  + b_3 \frac{\partial^2 \theta}{\partial r^2}
 + \frac{V_{\rm e} - V}{\tau} \, , \label{newvel}
\end{eqnarray}
and the variance equation
\begin{eqnarray}
 \frac{\partial \theta}{\partial t} + V \frac{\partial \theta}{\partial r} 
 &=& c_1 \frac{\partial \rho}{\partial r} 
 + c_2 \frac{\partial V}{\partial r}
 + c_3 \frac{\partial \theta}{\partial r} \nonumber \\
 &+& d_1 \frac{\partial^2 \rho}{\partial r^2} 
  + d_2 \frac{\partial^2 V}{\partial r^2}
  + d_3 \frac{\partial^2 \theta}{\partial r^2}
 + \frac{2(\theta_{\rm e} - \theta)}{\tau} \label{newvar}
\end{eqnarray}
(which corresponds to the equation of heat conduction in ordinary fluids).
Here, we have used the abbreviations
\begin{equation}\begin{array}{rclcrcl}
 a_1 &=& - \big[ \frac{1}{\rho} + (1-p) \chi s 
 (1 + \rho \chi l) \big]\theta \, , &\quad& 
 a_2 &=& (1-p) \chi \rho \Big(2s\sqrt{\frac{\theta}{\pi}} 
 - \rho \theta \chi \frac{l^2}{V} \Big) \, , \nonumber \\
 a_3 &=& - \big[ 1 + (1-p) \rho \chi \frac{s}{2} \big] \, , &\quad&
 b_1 &=& - (1-p) \chi s \Big( \frac{s}{2} + \rho \chi \frac{l^2}{2} \Big) 
 \theta \, , \nonumber \\
 b_2 &=& (1-p) \chi \rho \Big( s^2 \sqrt{\frac{\theta}{\pi}} 
 - \rho\theta\chi \frac{l^3}{2V} \Big)
 \, , &\quad&
 b_3 &=& - (1-p) \chi \rho \frac{s^2}{4} \, , \nonumber \\
 c_2 &=& - [2 + (1-p) \chi\rho s]\theta \vphantom{\frac{1}{2}} \, , &\quad&
 c_3 &=& 2(1-p) \chi \rho s \sqrt{\frac{\theta}{\pi}} \, , \nonumber \\
 d_2 &=& - (1-p) \chi \rho  s^2 \frac{\theta}{2} \, , &\quad&
 d_3 &=& (1-p) \chi \rho s^2 \sqrt{\frac{\theta}{\pi}} \, , 
\label{coeff}
\end{array}
\end{equation}
and
\begin{equation}
\begin{array}{rclcrcl}
 V_{\rm e} &=& V_0 - \tau (1-p) \rho \chi \theta \, , &\quad&
 \theta_{\rm e} &=& \tau D \, . \label{equi}
\end{array}
\end{equation}
Note that $c_1 = 0$ and $d_1 = 0$, which is a consequence of the assumed
interaction law of vehicles. 
It is one of the advantages of a kinetic derivation of macroscopic traffic
equations, that the above functions can be analytically calculated.
For example, we have obtained an expression for the equilibrium velocity $V_{\rm e}$. 
According to (\ref{equi}), it is given by the desired velocity $V_0$,
diminished by a term due to decelerating interactions. The latter
is proportional to the vehicle density and to the velocity variance, which
is very plausible. The function $\partial {\cal P}/\partial \rho 
:= - \rho \, a_1$ can be interpreted
as the partial derivative of the ``traffic pressure'' ${\cal P}$ with respect
to density. The quantity $\eta := \rho \, b_2$ has the
meaning of a viscosity, which smoothes out sudden spatial changes of the
velocity profile $V(r,t)$. Both quantities are non-negative and diverge
at maximum density $\rho_{\rm max}:= 1/l_0$, as it should be
for reasons of consistency \cite{VD,PhysA}. Previous macroscopic traffic models
did not describe these important facts correctly, since they
have neglected the terms in (\ref{coeff}) which explicitly contain
$l$ or $s$. Therefore, they are not valid for high vehicle densities, 
i.e. for congested conditions. Finally, note that it is possible to calculate
Navier-Stokes corrections of the coefficients $a_i$, $b_i$, $c_i$, and
$d_i$ \cite{Lett}.
\par
\begin{figure}[htbp]
\unitlength1.2cm
\begin{center}
\begin{picture}(11,4.6)
\put(-0.8,-0.9){\epsfig{width=11\unitlength, file=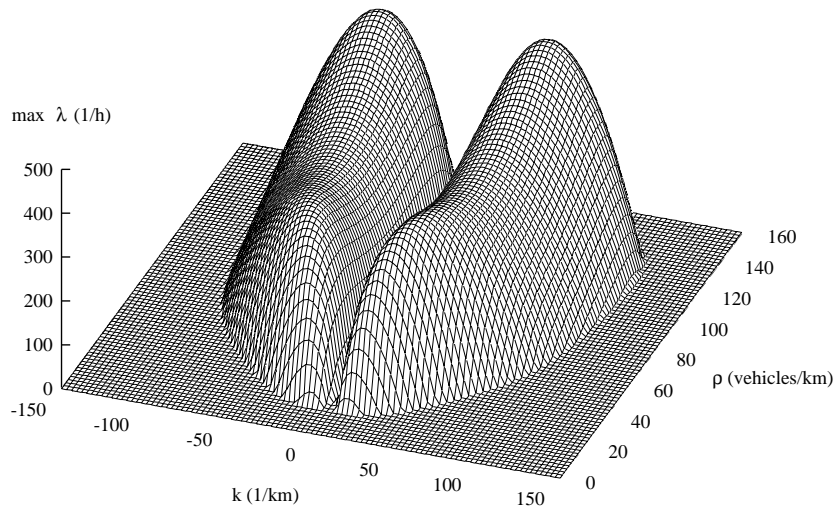}}
\end{picture}
\end{center}
\caption[]{Instability diagram for the Euler-like 
macroscopic traffic equations, including the dynamic variance equation  
\cite{Lett}.\label{F6}}
\end{figure}
According to our approximations,
equations (\ref{dens}) to (\ref{newvar}) are valid for small gradients
of $\rho$, $V$, and $\theta$. Therefore, they allow to investigate the
evolution of small disturbances of the (stationary and spatially homogeneous)
equilibrium solution. Figure \ref{F6} depicts the result of a linear
instability ana\-lysis, showing that traffic flow is only stable at
small and extreme densities as well as large wave numbers $|k|$
(i.e. small wave lengths $\ell = 2\pi/|k|$). This is in agreement with
empirical findings.
\par
\begin{figure}[h]
\unitlength1.2cm
\begin{center}
\begin{picture}(11,6.1)
\put(-0.55,-0.8){\epsfig{width=11\unitlength, file=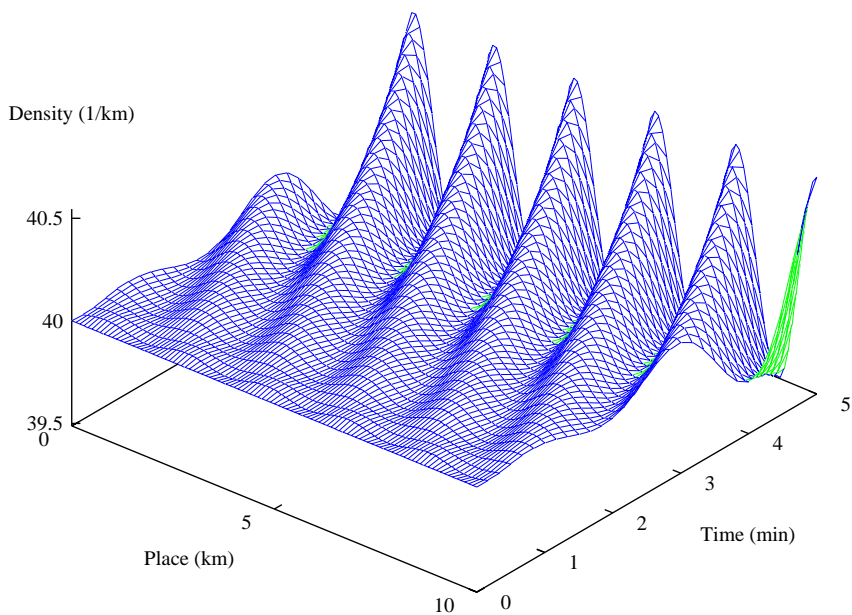}}
\end{picture}
\end{center}
\caption[]{Above a certain density, traffic flow is unstable. This
gives rise to the development of stop-and-go waves. In the represented
simulation, we
applied periodic boundary conditions (which corresponds to a circular 
road).\label{treiber}}
\end{figure}

The instability diagram is obtained by 
\begin{enumerate}
\item assuming a small
periodic perturbation
$\delta g(r,t) = g_0 \exp[ikr+(\lambda + {\rm i}\omega)t]$
of the macroscopic traffic quantities $g \in \{\rho, V, \theta\}$
relative to the stationary and spatially homogeneous 
equilibrium solution $g_{\rm e}(\rho)$ ($g_0$ being the amplitude,
$k$ the wave number, $\lambda$ the growth rate, and $\omega$ the frequency
of the perturbation), 
\item inserting $g(r,t)= g_{\rm e} + \delta g(r,t)$
into the macroscopic traffic equations,
\item neglecting quadratic terms in the small perturbations
$\delta g (r,t) \ll g_{\rm e}$, 
\item determining the three
complex eigenvalues $\tilde{\lambda} = \lambda + {\rm i} \omega$
of the linearized equations in dependence of 
$\rho$ and 
$k$. 
\end{enumerate}
An explicit example for this procedure is discussed in \cite{VD,PhysA}.
Equilibrium traffic flow is unstable if at least
one of the growth rates is positive, i.e. max $\lambda > 0$. Therefore,
the instability diagram shows max $\lambda(k,\rho)$ if this is greater
than zero, otherwise 0.
Figure \ref{treiber} depicts a simulation result
which demonstrates emerging stop-and-go traffic.

\subsection{Non-linear Phenomena}

As a consequence of the inherent non-linearity of the macroscopic traffic
equations, they display the self-organization of a number of collective
patterns of motion. This includes the formation of density clusters
(`traffic jams'), anti-clusters, dipole layers, 
cascades of density clusters (`stop-and-go
traffic'), and merging of density clusters. Moreover,
one finds subcritical instabilities and non-linear wave selection phenomena
\cite{KK1,KK2}.

\section{An Alternative Approach}

The Boltzmann-like formula (\ref{Boltz}) for vehicle interactions implicitly
assumes, that deceleration maneuvers happen instantaneously. This
approximation is only valid, if the average
duration of deceleration maneuvers is considerably smaller than the time
scale of the macroscopic traffic dynamics. However, one
can also derive fluid-dynamic traffic equations without this assumption.
We will illustrate this for a one-lane microscopic traffic model without
possibilities of overtaking. 

\subsection{A Concrete Microscopic Model}

\begin{figure}[htbp]
\unitlength6.5mm
\begin{center}
\begin{picture}(16,9.2)(0,0)
\put(0,9.8){\epsfig{height=16\unitlength, width=9.8\unitlength, angle=-90, 
      bbllx=50pt, bblly=50pt, bburx=554pt, bbury=770pt, 
      file=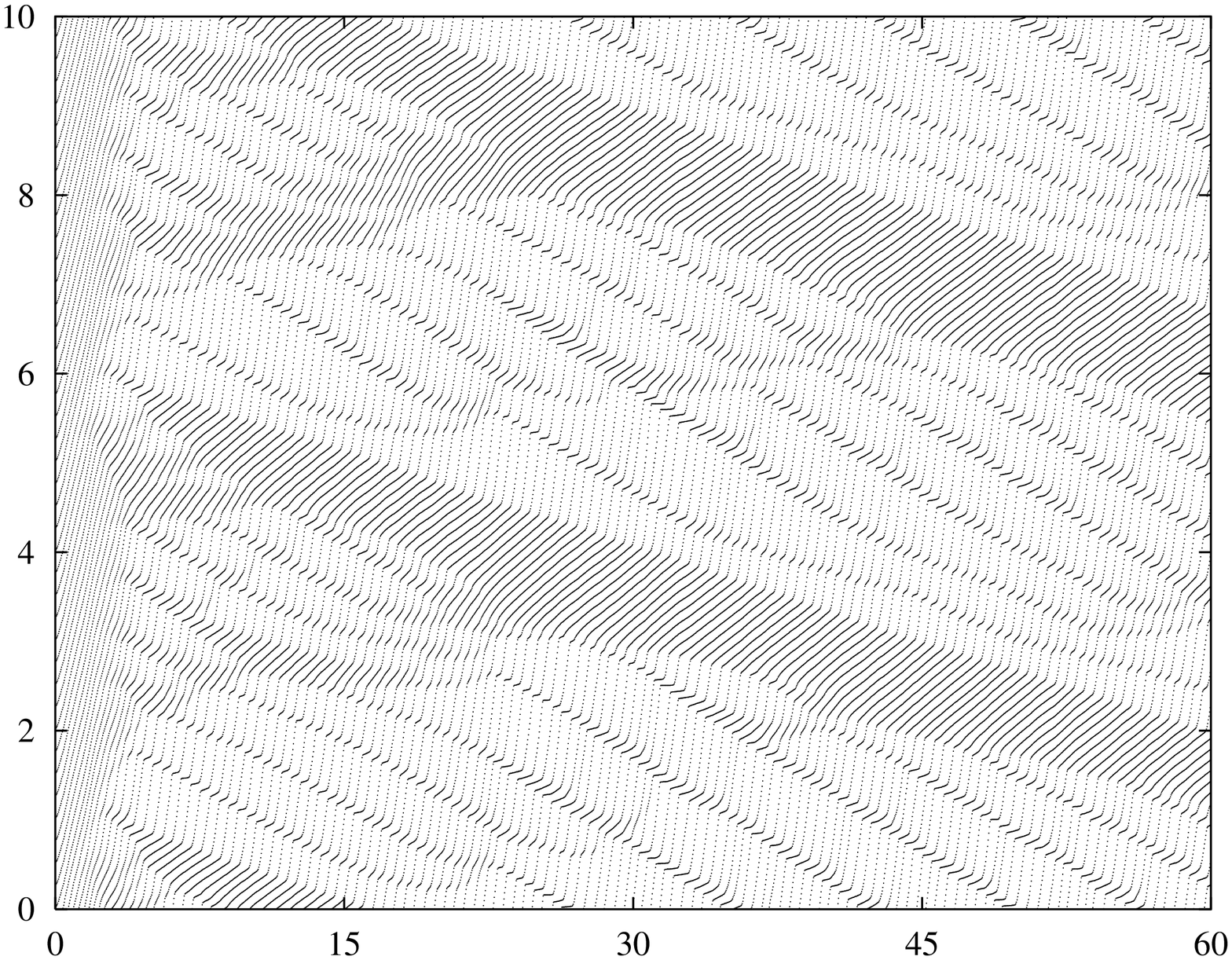}}
\put(8.7,-0.6){\makebox(0,0){\footnotesize $t$ (min)}}
\put(0.2,5.1){\makebox(0,0){\rotate[l]{\hbox{\footnotesize $r$ 
(km)}}}}
\end{picture}
\end{center}
\caption[]{Representation of the vehicle trajectories of each
10th car on a circular road \cite{VD}. The slopes of the trajectories indicate
the respective vehicle velocities, whereas their density reflects
the spatial vehicle density. The simulation starts with a homogeneous traffic 
situation (i.e. all vehicles have initially the same distance to the car in
front). 
In the course of time, density clusters emerge. These
are often called `phantom traffic jams', since they do not originate from
any localized bottleneck.\label{schwarz}}
\end{figure}

Let us start with the {\em social force model} of vehicle dynamics,
given by $dr_\alpha/dt = v_\alpha(t)$ and 
\begin{equation}
 \frac{dv_\alpha}{dt} = \underbrace{\frac{V_0 - v_\alpha(t)}
 {\tau}}_{\mbox{Acceleration}}
 + \underbrace{f_{\alpha (\alpha+1)}(r_\alpha,v_\alpha,r_{\alpha+1},
 v_{\alpha+1})}_{\mbox{Deceleration}}
 + \xi_\alpha(t) \, . 
\label{case}
\end{equation}
It is known that models of this kind are able to describe the
emergence of stop-and-go traffic \cite{VD,Bando} (cf. Figure \ref{schwarz}). 
\par
The advantage of the social force concept is, that it allows a very in\-tuitive
modelling of decision processes which are related to continuous changes in
some (possibly abstract) space of behavioral alternatives \cite{Quant}. 
According to this, the different motivations which influence an individual at
the same time, are described by additive,
force-like quantities. These generalized forces
are, of course, no Newtonian forces. For example, they do not obey the law
{\it actio = reactio}. The social force concept is well compatible with 
theoretical concepts from the social sciences and has been elaborated 
in detail \cite{Quant,VD}. It has already been successful
in describing various self-organization phenomena in 
pedestrian crowds \cite{VD,Peter},
but it was also applied to opinion formation processes \cite{Quant,PA}.
\par
In the case of driver behavior, we have two 
conflicting motivations: On the one hand,
the driver likes to accelerate towards his desired velocity $V_0$. On the
other hand, he wants to keep a safe distance to the car in front.
The latter is described by a repulsive deceleration force
$f_{\alpha(\alpha+1)}$. Effects $f_{\alpha\beta}$ of
interactions with other vehicles $\beta \ne (\alpha+1)$ have been assumed
to be negligible, here. However, they could easily be included in accordance
with Eq.~(\ref{force}). 
\par
\begin{figure}[htbp]
\unitlength=1cm
\begin{center}
\begin{picture}(10,7)
\put(-2,-0.3){\epsfig{width=12\unitlength, file=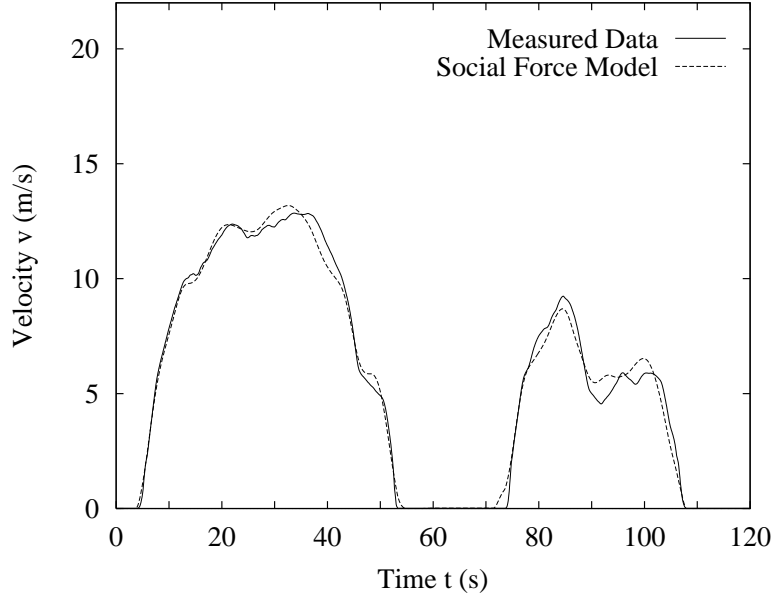}}
\end{picture}
\end{center}
\caption[]{Time-dependent velocity 
of a car which follows another car in city traffic. The simulation
results for the social force model 
treat the velocity of the vehicle ahead and its initial distance
as given. We find a good agreement with the empirical follow-the-leader
data.\label{benno}}
\end{figure}

As Fig.~\ref{benno} shows, a good agreement with empirical data of 
driver-vehicle behavior can be achieved with the following form of
the repulsive interaction force:
\begin{equation}
 f_{\alpha(\alpha+1)} := \frac{V'_{\rm e}(r_{\alpha+1} - r_\alpha) 
 - V_0}{\tau} + f'_{\alpha(\alpha+1)} 
\label{fab}
\end{equation}
with
\begin{equation}
 f'_{\alpha(\alpha+1)} := - \exp\left(-\frac{r_{\alpha+1} - r_\alpha - s(v_\alpha)}{R} \right)
 \frac{v_\alpha - v_{\alpha+1}}{\tau'} \, \Theta(v_\alpha - v_{\alpha+1}) \, .
\label{Fab}
\end{equation}
Here, $\Theta(\Delta v)$ is the Heaviside step function.
\par
If we would restrict the model to the first term of (\ref{fab}) 
(i.e. in the case
$\tau' \rightarrow \infty$), we would arrive at the microsimulation model
by Bando et al. \cite{Bando}. $V'_{\rm e}(\Delta r)$ is the equilibrium
velocity, which is a function of the distance $\Delta r := r_{\alpha+1}
- r_\alpha$ to the next car ahead. The additional term 
(\ref{Fab}) takes into account that
\begin{enumerate}
\item drivers brake stronger, when the relative velocity
$\Delta v := v_\alpha - v_{\alpha+1}$ is large or when the
distance $\Delta r$ to the car in front is small,
\item the deceleration time $\tau'$ is shorter than the acceleration time
$\tau$,
\item drivers begin to brake at a larger distance, if they drive fast.
This is described by the velocity-dependence of the safe distance
$s(v)$. $R$ is the range of the repulsive effect of a car.  
\end{enumerate} 
In most microsimulations, the relations 
$V'_{\rm e}(\Delta r)$ and parameters $\tau$, $\tau'$, $R$ are
specified individually (i.e. in an $\alpha$-dependent way).
\par
It can be shown that the above force model is consistent in the
limiting cases. For large distances or $v_\alpha \approx v_{\alpha+1}$,
vehicle $\alpha$ approaches the distance-dependent equilibrium velocity
$V'_{\rm e}$:
\begin{equation}
 \frac{dv_\alpha}{dt} \approx \frac{V'_{\rm e}(r_{\alpha+1} - r_\alpha) 
 - v_\alpha(t)} {\tau} + \xi_\alpha(t) \, .
\end{equation}
For small distances and $v_\alpha > v_{\alpha+1}$, it decelerates to the
velocity $v_{\alpha+1}$ of the car in front: 
\begin{equation}
 \frac{dv_\alpha}{dt} \approx \frac{v_{\alpha+1}(t) - v_\alpha(t)}
 {\tau' \mbox{e}^{[r_{\alpha+1} - r_\alpha - s(v_\alpha)]/R} }
 + \xi_\alpha(t) \, .
\end{equation}
With decreasing distance it brakes stronger.

\subsection{Derivation of Macroscopic Traffic Equations}

In the following, we write the acceleration relation in the form
\begin{equation}
 \frac{dv}{dt} = f(\Delta r,v,w) + \xi(t)
 := \frac{V'_{\rm e}(\Delta r) - v}{\tau} 
 + f'(\Delta r, v, w) + \xi(t)
\label{eff}
\end{equation}
with the abbreviations $v := v_\alpha$,
$w := v_{\alpha+1}$, 
$\Delta r := r_{\alpha+1} - r_\alpha$, $\xi := \xi_\alpha$,
and
\begin{equation}
 f' := f'_{\alpha(\alpha+1)} = f_{\alpha(\alpha+1)} 
 + \frac{V_0 - V'_{\rm e}(\Delta r)}{\tau} \, .
\end{equation}
Relation (\ref{eff}) is now inserted into the kinetic equation
\begin{equation}
 \frac{\partial \tilde{\rho}}{\partial t}
 + \frac{\partial (\tilde{\rho} v)}{\partial r}
 + \frac{\partial (\tilde{\rho} f)}{\partial v}  
 = \left( \frac{\partial \tilde{\rho}}{\partial t} \right)_{\rm fl}
 \, ,
\end{equation}
which is again a direct consequence of definition (\ref{phase}).
Note that the interaction effects were absorbed into the function $f$, here.
Next, we multiply this equation with $v^k$ and $P'(w,\Delta r|v,r,t)$,
which denotes the probability that, given a car with velocity $v$ is located
at place $r$, the car in front drives with velocity $w$ at a distance $\Delta r$.  
Finally, the resulting equation is integrated with respect to $w$ and
$\Delta r$. This gives the macroscopic equations
\begin{equation}
 \frac{\partial \rho}{\partial t} + \frac{\partial (\rho V)}{\partial r}
 = 0 \, ,
\label{ncont}
\end{equation}
\begin{equation}
 \frac{\partial (\rho V)}{\partial t} + \frac{\partial}{\partial r}
 [\rho (V^2 + \theta)] = \frac{\rho}{\tau} ( V_{\rm e}^* - V ) 
 + \rho {\cal F}_1 \, ,
\label{nvel}
\end{equation}
\begin{equation}
 \frac{\partial}{\partial t} [\rho (V^2 + \theta)] +
 \frac{\partial}{\partial r} [\rho (V^3 + 3 V \theta )]
 = \frac{2\rho}{\tau} (V_{\rm e}^*V + \tau D - V^2 - \theta)
 + \rho {\cal F}_2 
\label{nvar}
\end{equation}
with
\begin{eqnarray}
 {\cal F}_k (r,t) &:=& k \int d\Delta r \int dv \int dw \, v^{k-1}
 f'(\Delta r,v,w) P'(\Delta r,w|r,v,t) \frac{\tilde{\rho}(r,v,t)}{\rho(r,t)} 
 \nonumber \\
 &=& - k \int d\Delta r \int dv \int\limits_{w < v} dw \, v^{k-1}
 \exp\left(-\frac{\Delta r - s(v)}{R} \right)
 \frac{v-w}{\tau'} \nonumber \\
 & & \times P'(\Delta r,w|r,v,t) P(v;r,t)
\label{zw}
\end{eqnarray}
and
\begin{equation}
 V_{\rm e}^* := \int d\Delta r \int dv \int dw \, V'_{\rm e}(\Delta r)
 P'(\Delta r,w|r,v,t) P(v;r,t) \, .
\end{equation}
To get (\ref{zw}), we made use of partial integration.
\par
In the following, we will apply the factorization approximation
\begin{equation}
 P'(\Delta r, w|r,v,t) \approx P_*(\Delta r;r,t) P(w;r+\Delta r,t) \, ,
\end{equation}
which is even exact, if the distributions of the 
velocities $w$ and the headways $\Delta r$ are
statistically independent, and independent of $v$. Furthermore, we 
assume that the headway distribution $P_*(\Delta r;r,t)$ 
is a function of the density $\rho$ and average velocity $V$ 
at a certain place $r+\delta r$ between $r$ and $r+\Delta r$:
\begin{equation}
 P_*(\Delta r;r,t) \equiv P_*\Big(\Delta r;\rho(r+\delta r,t),
 V(r+\delta r,t)\Big) \, .
\label{this}
\end{equation} 
Then, we can expand (\ref{zw}) and (\ref{this}) in the small quantities
$\delta v := v - V$ and $\delta r$, respectively. In this way,
we obtain
\begin{eqnarray}
 {\cal F}_k (r,t) 
 &\approx& - k \int d\Delta r \int dv \int\limits_{w < v} dw \, v^{k-1}
 \frac{v-w}{\tau'} \exp\left(-\frac{\Delta r - s(V)}{R} \right) 
 \bigg\{ \frac{ds}{dV} \frac{\delta v}{R} \nonumber \\
 & & + \frac{1}{2} \bigg[ \frac{d^2 s}{dV^2} 
 + \bigg(\frac{ds}{dV} \bigg)^{\!2} 
 \bigg] \frac{\delta v^2}{R^2} \bigg\}
 P_*(\Delta r;r,t) P(w;r+\Delta r,t) P(v;r,t)
\nonumber \\
 & & 
\end{eqnarray}
and
\begin{eqnarray}
 P_*(\Delta r;r,t) \approx P_*\Big(\Delta r;\rho(r,t),V(r,t)\Big)
 &+& \frac{\partial P_*}{\partial \rho} \left(
 \frac{\partial \rho}{\partial r} \delta r + 
 \frac{\partial^2 \rho}{\partial r^2} \frac{\delta r^2}{2} \right)
 \nonumber \\
 &+& \frac{\partial P_*}{\partial V} \left( \frac{\partial V}{\partial r}
 \delta r + \frac{\partial^2 V}{\partial r^2} \frac{\delta r^2}{2}
 \right) \, , \qquad
\end{eqnarray}
where we again neglected products of partial derivatives $\partial g/\partial
r$. After carrying out the integrations over $v$, $w$, and $\Delta r$,
the resulting macroscopic traffic equations can again be written in the
form of Eqs. (\ref{dens}) to (\ref{newvar}). However, the 
coefficients $a_i$, $b_i$, $c_i$, and $d_i$ are different,
since we did not apply the approximation of sudden deceleration maneuvers.  
The problem of this method is, that it does not provide the functional
form of the headway distribution $P_*(\Delta r;\rho,V)$, which is needed
for the explicit evaluation of the coefficients. Nevertheless, the
use of the above results will be presented by a simple example.

\subsection{Relation between Bando's microscopic and Payne's macroscopic
traffic model}

The microsimulation model by Bando
et al. \cite{Bando} is obtained by neglecting the fluctuation term and
the second term in (\ref{fab}), i.e. by setting $D:= 0$ and
$f' := 0$. In order to calculate the corresponding macroscopic traffic
equations, we make a very simple assumption, here, namely that the
headways $\Delta r$ are given by the inverse of the density:
\begin{equation}
 P_*(\Delta r;r,t) := \delta \left( \Delta r - \frac{1}{\rho(r+\delta r,t)}
 \right) \, .
\end{equation}
Inserting this into the above equations, we finally arrive at
the continuity equation
\begin{equation}
 \frac{\partial \rho}{\partial t} + V \frac{\partial \rho}{\partial r}
 = - \rho \frac{\partial V}{\partial r} \, ,
\label{one}
\end{equation}
the velocity equation
\begin{eqnarray}
 \frac{\partial V}{\partial t} + V \frac{\partial V}{\partial r}
 &\approx & - \frac{1}{\rho} \frac{\partial (\rho \theta)}{\partial r}
 + \frac{1}{\tau} \!\left[ V_{\rm e}^*\!\left(\frac{1}{\rho}\right)\!
 - V \right] \!
 - \frac{1}{\tau \rho^2} \frac{\partial V_{\rm e}^*}{\partial \Delta r}
 \left( \frac{\partial \rho}{\partial r} \delta r
 + \frac{\partial^2 \rho}{\partial r^2} \frac{\delta r^2}{2} \right)
 \nonumber \\  
 &\approx & - \frac{1}{\rho} \frac{\partial (\rho \theta)}{\partial r}
 + \frac{1}{\tau} [ V_{\rm e} (\rho ) - V ]
 + \frac{1}{\tau} \frac{\partial V_{\rm e}}{\partial \rho}
 \left( \frac{\partial \rho}{\partial r} \delta r
 + \frac{\partial^2 \rho}{\partial r^2} \frac{\delta r^2}{2} \right)\, ,   
\label{two}
\end{eqnarray}
and the variance equation
\begin{equation}
 \frac{\partial \theta}{\partial t} + V \frac{\partial \theta}{\partial r}
 = -2\theta \frac{\partial V}{\partial r} - \frac{2}{\tau} \theta \, ,
\end{equation}
where
\begin{equation}
 V_{\rm e}(\rho) := V_{\rm e}^*\left(\frac{1}{\rho}\right) \, .
\end{equation}
Close to the equilibrium solution, the variance equation can be neglected
due to $\theta \approx 0$.
The instability condition of the
remaining equations (\ref{one}) and (\ref{two}) reads
\begin{equation}
 \rho \left| \frac{dV_{\rm e}}{d\rho} \right| 
 \stackrel{!}{>} \frac{\delta r}{\tau}
\end{equation}
(cf. \cite{VD,PhysA}).
This is only compatible with the instability condition
\begin{equation}
 \frac{dV_{\rm e}^*}{d\Delta r} \stackrel{!}{>} \frac{1}{2\tau}
 \qquad \mbox{or} \qquad 
 \rho^2 \left| \frac{dV_{\rm e}}{d\rho} \right|
 \stackrel{!}{>} \frac{1}{2\tau} 
\end{equation}
of the Bando model \cite{Bando}, if we choose 
\begin{equation}
  \delta r \stackrel{!}{=} \frac{1}{2\rho} \approx \frac{\Delta r}{2} \, ,
\end{equation}
which is very plausible. In this case, the macroscopic equations
(\ref{one}) and (\ref{two}) agree with the traffic model by Payne \cite{Payne},
but they contain the
additional term $[\delta r^2/(2\tau)] (\partial V_{\rm e}/\partial \rho)
  \partial^2 \rho/\partial r^2$, which describes a smoothing of sudden spatial
  changes in density and velocity. However, as soon as
the approximation $\theta \approx 0$ becomes invalid, Payne's model does
not anymore reflect the traffic dynamics according to Bando's model. 

\section{Summary and Conclusions}

We have presented microscopic and macroscopic traffic flow models for
freeways, which were successfully confronted with empirical data
(cf. also \cite{VD,PhysA,empi}). Moreover, it has been shown, how
macroscopic traffic models can be systematically derived from the
equations of motion for single vehicles. This is essential for increasing
the speed of traffic simulations. Apart from the kinetic approach
to this problem, which based on the assumption of 
sudden deceleration maneuvers, an
alternative method has been proposed, which presupposes a suitable
approximation of the headway distribution function. The resulting
macroscopic traffic equations are related to the hydrodynamic equations
of ordinary fluids, but they contain a number of additional terms for three
reasons: 
\begin{enumerate}
\item Vehicles accelerate to a certain desired velocity.
\item A finite equilibrium variance of vehicle velocities is caused by
imperfect driving.
\item Vehicle interactions are anisotropic and do not conserve energy
or momentum.
\end{enumerate}
The additional terms are responsible for certain instabilities of
traffic flow, causing `phantom traffic jams' or `stop-and-go traffic'.
They are also the origin of viscosity effects and of the divergence of
`traffic pressure' at maximum vehicle density. Here, it is essential
that vehicular space requirements are taken into account \cite{VD,PhysA,Lett}.
Otherwise, the macroscopic traffic model would neglect certain characteristic
terms, which would limit its validity to non-congested traffic situations. 
\par
For the purpose of computer simulations, it is advantageous to have
the macroscopic traffic equations in flux representation. This has
been ana\-lytically derived, but it contains the Gaussian error function.
In contrast to previous results \cite{Lett},
the corresponding equations are not restricted to cases of small gradients.

\section*{Acknowledgments}
The author wants to thank Martin Treiber, Tilo Schwarz,
and Benno Tilch for providing
Figs. \ref{treiber}, \ref{schwarz}, and \ref{benno}, respectively. 
He is also grateful to Henk Taale and
the Dutch Ministry of Transport, Public
Works and Water Management as well as 
to Thomas Bleile and the Robert Bosch GmbH for providing the
empirical data shown in Figs. 
\ref{inst} to \ref{F4}, and \ref{benno}, respectively.
The presented work has been financially supported by
the DFG, Heisenberg scholarship He 2789/1-1, and by the BMBF, grant no.
13N7092 (collaborative research project ``SANDY'').
%
%


\begin{thebibliography}
%
%
\bibitem{[1]}{follow1}{[1]} Gazis\ D.C., Herman,\ R., Rothery\ R.W. (1961):
Nonlinear Follow the Leader Models of Traffic Flow. Operations
Research {\bf 9}, 545--567

\bibitem{[2]}{follow2}{[2]} May\ A.D., Jr., Keller\ H.E.M. (1967): Non-Integer
Car-Following Models. Highway Research Record {\bf 199}, 19--32

\bibitem{[3]}{Wiede}{[3]} Wiedemann\ R. (1974): {\it Simulation des
    Stra{\ss}enverkehrsflusses} (Heft 8 der Schriftenreihe des IfV,
Institut f\"ur Verkehrswesen, Universit\"at Karlsruhe)

\bibitem{[4]}{Bando}{[4]} Bando\ M., Hasebe\ K., Nakayama\ A., Shibata\ A.,
Sugiyama\ Y. (1995): Dynamical Model of Traffic Congestion and Numerical
Simulation. Phys. Rev. E {\bf 51}, 1035--1042

\bibitem{[5]}{VD}{[5]} Helbing\ D. (1997): {\it Verkehrsdynamik: Neue physikalische
Modellierungskonzepte} (Springer, Berlin)

\bibitem{[6]}{Cell1}{[6]} Schreckenberg\ M., Schadschneider\ A., Nagel\ K., Ito\ N. 
(1996): Discrete Stochastic Models for Traffic Flow. Phys. Rev. E
{\bf 51}, 2939--2949.

\bibitem{[7]}{Cell2}{[7]} Nagel\ K., Paczuski\ M. (1995): Emergent Traffic Jams.
Phys. Rev. E {\bf 51}, 2909--2918

\bibitem{[8]}{Cell3}{[8]} Nagatani\ T. (1995): Bunching of Cars in Asymmetric
  Exclusion Models for Freeway Traffic. Phys. Rev. E {\bf 51}, 922--928

\bibitem{[9]}{Cell4}{[9]} Krauss\ S., Wagner\ P., Gawron\ C. (1996): Continuous
Limit of the Nagel-Schreckenberg Model. Phys. Rev. E {\bf 54}, 3707--3712

\bibitem{[10]}{Payne}{[10]} Payne\ H.J. (1971): Models of Freeway Traffic
and Control. In: Bekey\ G.A. (ed.) {\it Mathematical Models of Public Systems,
Vol.~1} (Simulation Council, La Jolla, CA), 51--61

\bibitem{[11]}{KK1}{[11]} Kerner\ B.S., Konh\"auser\ P. (1994): Structure and
  Parameters of Clusters in Traffic Flow. Phys. Rev. E {\bf 50}, 54--83

\bibitem{[12]}{KK2}{[12]} Kerner\ B.S., Konh\"auser\ P., Schilke\ M. (1995):
Deterministic Spontaneous Appearance of Traffic Jams in Slightly Inhomogeneous
Traffic Flow. Phys. Rev. E {\bf 51}, 6243--6246

\bibitem{[13]}{PhysA}{[13]} Helbing\ D. (1996): Derivation and Empirical Validation
of a Refined Traffic Flow Model. Physica A {\bf 233}, 253--282

\bibitem{[14]}{empi}{[14]} Helbing\ D. (1997): Empirical Traffic Data and
their Implications for Traffic Modeling. Phys. Rev. E {\bf 55}, R25--R28

\bibitem{[15]}{Kin1}{[15]} Prigogine\ I., Herman\ R. (1971): Kinetic Theory of
Vehicular Traffic (Elsevier, Amsterdam)

\bibitem{[16]}{Kin2}{[16]} Paveri-Fontana\ S.L. (1975): On Boltzmann-like Treatments
  for Traffic Flow. A Critical Review of the Basic Model and an Alternative
Proposal for Dilute Traffic Analysis. Transportation Research {\bf 9},
225--235

\bibitem{[17]}{Kin3}{[17]} Helbing\ D. (1996): Gas-Kinetic Derivation of
  Navier-Stokes-Like Traffic Equations. Phys. Rev. E {\bf 53}, 2366--2381

\bibitem{[18]}{Kin4}{[18]} Wagner\ C., Hoffmann\ C., Sollacher\ R., Wagenhuber\
  J., Sch\"urmann\ B. (1996): Second Order Continuum Traffic
  Flow Model. Phys. Rev. E {\bf 54}, 5073--5085

\bibitem{[19]}{Lanes1}{[19]} Helbing\ D., Greiner\ A. (1997): Modelling and
Simulation of Multilane Traffic Flow. Phys. Rev. E {\bf 55}, 5498--5508

\bibitem{[20]}{Lanes2}{[20]} Helbing\ D. (1997): Modeling Multi-Lane Traffic Flow
with Queuing Effects. Physica A {\bf 242}, 175--194

\bibitem{[21]}{chapman}{[21]} Chapman\ S., Cowling\ T.G. (1970): {\it
The Mathematical Theory of Nonuniform Gases} (3rd edition, Cambridge
University Press, Cambridge)

\bibitem{[22]}{densegran1}{[22]} Jenkins\ J.T., Richman\ M.W. (1985): Kinetic
Theory for Plane Flows of a Dense Gas of Identical, Rough, Inelastic, Circular
Disks. Phys. of Fluids {\bf 28}, 3485--3494 

\bibitem{[23]}{densegran2}{[23]} Lun\ C.K.K., Savage\ S.B., Jeffrey\ D.J.,
Chepurniy\ N. (1984): Kinetic Theories for Granular Flow: Inelastic Particles
in Couette Flow and Slightly Inelastic Particles in a General Flowfield.
J. Fluid. Mech. {\bf 140}, 223--256

\bibitem{[24]}{gold}{[24]} Goldshtein\ A., Shapiro\ M. (1995):
Mechanics of Collisional Motion of Granular Materials. Part 1.
General Hydrodynamic Equations. J. Fluid. Mech. {\bf 282}, 75--114

\bibitem{[25]}{emp}{[25]} Helbing\ D. (1997): Fundamentals of Traffic Flow.
Phys. Rev. E {\bf 55}, 3735--3738

\bibitem{[26]}{Lett}{[26]} Helbing\ D. (1997): Structure and Instability of Consistent 
High-Density Equations for Traffic Flow. Phys. Rev. Lett., submitted

\bibitem{[27]}{Quant}{[27]} Helbing\ D. (1995): {\it Quantitative Sociodynamics.
Stochastic Methods and Models of Social Interaction Processes}
(Kluwer Academic, Dordrecht)

\bibitem{[28]}{Peter}{[28]} Helbing\ D., Moln\'{a}r\ P. (1995): Social
Force Model for Pedestrian Dynamics. Phys. Rev. E {\bf 51}, 4282--4286

\bibitem{[29]}{PA}{[29]} Helbing\ D. (1993): Boltzmann-like and
  Boltzmann-Fokker-Planck Equations as a Foundation of Behavioral Models.
Physica A {\bf 196}, 546--573

\bibitem{}{}{}\par
For further references cf. \cite{VD}.
\end{thebibliography}
\end{document}